\documentclass[11pt]{amsart}
\pdfoutput=1

\usepackage[utf8]{inputenc}

\usepackage[margin=3.25cm]{geometry}

\usepackage{amssymb, amsmath, amsthm, amsfonts}
\usepackage{graphicx}
\usepackage{color}
\usepackage{mathrsfs}
\usepackage{physics}
\usepackage{mathtools}
\usepackage{tikz}
\usepackage{faktor}
\usepackage{tikz-cd}
\usepackage{bm}
\usepackage[hidelinks]{hyperref}
\usepackage{csquotes}

\DeclareRobustCommand{\SkipTocEntry}[5]{}

\newtheorem{theorem}{Theorem}
\newtheorem{lemma}{Lemma}
\newtheorem{corollary}{Corollary}
\newtheorem{defn}{Definition}
\newtheorem{prop}{Proposition}

\newtheorem{conjecture}{Conjecture}
\theoremstyle{remark}
\newtheorem{remark}{Remark}
\theoremstyle{remark}

\theoremstyle{remark}
\newtheorem*{notation*}{Notation}

\usepackage[UKenglish]{isodate}
\usepackage[UKenglish]{babel}

\newcommand{\R}{\mathbb{R}}

\DeclareMathOperator{\E}{\mathbb{E}}

\renewcommand{\grad}{\nabla}

\newcommand{\Ind}{\mathrm{Ind}}
\newcommand{\nInd}{{\mathrm{nInd}}}

\title[Weak Markov blankets]{Weak Markov blankets in high-dimensional, sparsely-coupled random dynamical systems}

\author{Dalton A R Sakthivadivel}
\address{\parbox{\linewidth-12pt}{Department of Mathematics, Department of Physics and Astronomy, Stony Brook University, Stony Brook, NY, USA, 11794-3651}}
\address{VERSES Research Lab and Spatial Web Foundation, Los Angeles, CA, USA, 90016}
\email{dalton.sakthivadivel@stonybrook.edu}
\urladdr{https://darsakthi.github.io}

\date{\today}

\let\oldtocsection=\tocsection

\let\oldtocsubsection=\tocsubsection

\renewcommand{\tocsection}[2]{\hspace{0em}\oldtocsection{#1}{#2}}
\renewcommand{\tocsubsection}[2]{\hspace{19.5pt}\oldtocsubsection{#1}{#2}}

\begin{document}

\maketitle

\begin{abstract}

This paper formulates a notion of high-dimensional random dynamical systems that couple to another system, like an embedding environment, in such a way that each system engages in controlled exchange with the other system. Using the adiabatic theorem and asymptotic arguments about the behaviour of systems with many degrees of freedom, we show that this sort of controlled, but not strictly sparse, coupling structure is ubiquitous in high-dimensional systems. In doing so we prove a conjecture of K Friston.

\end{abstract}

%%%%%%%%%%%%%%%%%%%%%%%%%%%%%%%%%%%%%%%%%%%%%%%%%%%%%%%%%%%%%%%%%%%%%

\section{Introduction}

Let a random dynamical system on $\R^n$ satisfy some base flow on a probability space over $\R^n$ and some additive deterministic flow on $\R^n$ \cite{arnoldRDS}, and let such a system be coupled to another system if some subset of each system's state variables appear in the deterministic component of the flow of the other respective system. When coupled, we must ask what prevents these systems from mixing with one another such that their statistical properties become the same. Should the two systems mix, we could not sample from $\eta$ and conclude that it is $\eta$ as opposed to $\mu$: any such assignment would be vacuous. Understanding the structure of such systems is an interesting question from the point of view of statistical physics and the study of stochastic dynamical systems.

As introduced in literature on complex and biophysical systems\textemdash which consists essentially of the study of various sorts of coupling structures in random dynamics and the dynamical properties they yield\textemdash the so-called \emph{Markov blanket} is what encodes the distinguishability of the statistics of one system from another \cite{friston12, afepfapp}. 

\begin{defn} Let $\eta$ and $\mu$ be two random dynamical systems and let the expectations of $\eta$ and $\mu$ be sufficient statistic for these systems, uniquely parameterising the probability measure over the state space. A Markov blanket $b$ is a coupling state between $\eta$ and $\mu$ such that the statistics of $\eta$ and $\mu$ are independent when conditioned on $b$, i.e.,
\begin{equation}\label{mb-cond}
\E_{p(\eta \mid b, \eta)}[\eta \mid b, \mu] = \E_{p(\eta \mid b)}[\eta \mid b]
\end{equation}
and vice-versa for $\mu$.
\end{defn}

This terminology is owed to the fact that, when conditioned on blanket states, any such set of systems is Markovian (this is apparent from \eqref{mb-cond}, a conditional independence in $\eta$ and $\mu$). It originates in the theory of Markov chains and their graphical structures \cite{pearl}.

Current challenges to any realistic modelling of biological physics include the fact that complex biophysical systems engage in dynamic, fluctuating interactions with an environment, and are not perfectly stationary nor sequestered from their environment, despite being controlled.\footnote{The author thanks both Ken A Dill and Maxwell J D Ramstead for highlighting this point in conversation.} This is not to mention the physical fact that perfectly isolated systems are difficult or impossible to realise. Random dynamical systems equipped with Markov blankets are apt representations of the physics of exactly these situations, making them interesting objects to study, but initially there appear to be strong conditions on the existence of such systems\textemdash blanketed systems are far more separable than most physical systems ought to be. In this vein, previous literature has emphasised the question of whether Markov blankets can actually be found in faithful models of physical systems \cite{aguilera2021}, whilst other work has led to the conjecture that they are, in fact, common in sufficiently complex systems \cite{stoch-chaos}. This latter statement is referred to as the \emph{sparse coupling conjecture}.

Here we define a notion of a weak Markov blanket that accommodates greater exchange without sacrificing conventional notions of separability or physical control. We go on to show that this notion of a blanket is, in fact, ubiquitous in high-dimensional random dynamical systems with multiple timescales, in the sense that a sufficient condition for the presence of such a weak Markov blanket is met with probability one asymptotically with system dimension. Our approach will be as follows: we will define a class of systems with an initial type of sparsity; this furnishes these systems with a mathematical structure which is useful for modelling systems in statistical physics. Proving the sparse coupling conjecture asymptotically is possible on this set, such that counterexamples to the sparse coupling conjecture exist in a set of measure zero for generic physical systems. This achieves a weakened statement of the original conjecture, in such a way that a good result can be obtained which is still physically sensible.

\subsection*{Acknowledgements} The author thanks Karl J Friston,  Lancelot Da Costa, and Conor Heins for invaluable discussions about the sparse coupling conjecture.

\section{Markov blankets; stronger, weaker, middling}

A Markov blanket is what allows us to read off the separability of two interacting systems. The presence of a Markov blanket is equivalent to a particular sort of control, that of fixed points of the system via the maintenance of a low-surprisal attractor in the state space, in the sense of \cite[Theorem 4.2]{g-and-a}. At a high level, we can sketch out a proof by contradiction to demonstrate that any two coupled random dynamical systems are necessarily blanketed, and conversely, that a blanket is sufficient to distinguish two random dynamical systems. The latter actually follows by construction. Hence we will focus on the former statement. Suppose we have two random dynamical systems $\eta$ and $\mu$. Now we assume their statistical properties are indistinguishable. Consequently, their samples will be identical. This means the systems mix in such a way that they are indistinguishable. Therefore, by contradiction, there must be a blanket distinguishing one system from the other; if not, the entire formalism degenerates into a single system. In other words, a system which has the same statistics as another system cannot be a statistically distinct system; we make the further claim that a system with the same statistics as another system cannot be a physically distinct system based on elementary notions from statistical physics about ergodic sampling. The aim of this paper is to establish this proof in greater detail: that a Markov blanket necessarily exists for a large class of physically-relevant coupled but separated systems.

Suppose $\eta$ (likewise for $\mu$) satisfies some It\=o stochastic differential equation and admits a steady state density $p(\eta \mid b)$. The general form (see \cite{afepfapp, barp}) of such a measure-preserving diffusion is given by 
\begin{equation}\label{sde}
\dd{\eta_t} = -(\Gamma - Q)\grad U(\eta, b, \mu)\dd{t} + \sqrt{2D}\dd{W_t}
\end{equation}
for some skew-symmetric matrix $Q$, positive definite $\Gamma$, and scalar function 
\[
U : \R^k \cross \R^l \cross \R^m \to \R, \qquad k+l +m = n.
\]
Since $Q$ is skew-symmetric, it has exclusively non-zero off-diagonal entries, meaning the relationships between different state variables are encoded in $Q$. That is to say, the influence of one variable on another is encoded in the mixing of flow directions in state space due to $Q$. 

We assume that $U$ is a logarithmic function of states, known as the surprisal in statistics and information geometry. Given a probability measure $\exp{-J(\eta, b, \mu)}$ with $J$ some scalar function, $U$ is equivalent to $J$. We typically assume $J$ is the quadratic function $(\mu - \sigma^{-1}(\hat\eta_b))^2,$ as described in \cite{dacosta, classical-physics}, and as such, the Hessian is constant, and the steady state solution to \eqref{sde} is a Gaussian.\footnote{We describe the general use of quadratic surprisals as approximations to non-equilibria in \cite[Section 6.3]{g-and-a}. We will generalise past this assumption in Sections \ref{quadratic-section} and \ref{non-lin-section} (cf. Remark \ref{gaussian-remark}).} The nature of the partial function $\xi(b, -)(\mu) = \sigma(\hat\mu_b) = \hat\eta_b$ is described in \cite{g-and-a}, creating some set-point for the $\mu$ system based on the way $\eta$ is likely behaving given the blanket states coupling the two. We also rename the matrix operator $\Gamma - Q$ to $T$. Couplings between different states are in some sense weighted by $T$. Since the Jacobian of this flow is seen to be 
\[
J = T \wedge \partial^2 U + \grad T \wedge \grad U + \grad (\grad \cdot T),
\]
or 
\begin{equation}\label{matrix-mult}
J_{ij} = \sum T_{it} \partial_{tj} U + \sum \partial_j T_{it} \partial_t U + \partial_{jt} T_{it}
\end{equation}
in components, where $\partial_{ij} U$ denotes the $ij$-th entry of the Hessian form $\partial^2 U$, we have
\begin{equation}\label{jacobian}
\begin{bmatrix}
J_{\eta^1\eta^1} &\ldots &J_{\eta^1 b^1} &\ldots &J_{\eta^1\mu^1} &\ldots &J_{\eta^1\mu^m}\\
\vdots &\ddots &\vdots\\
J_{\eta^k\eta^1} &\ldots & J_{\eta^k b^1} &\ldots\\
J_{b^1 \eta^1} &\ldots & J_{b^1 b^1 } &\ldots &J_{b^1 \mu^1} &\ldots \\
\vdots &\ddots &\vdots \\
J_{\mu^1\eta^1} &\ldots &J_{\mu^1\mu^1} &\ldots\\
\vdots & & & & & \ddots \\
J_{\mu^m \eta^1} &\ldots & & & & & J_{\mu^m\mu^m}
\end{bmatrix}
\end{equation}
in the sense that the flow in any state space direction depends on its interactions with flows in other directions.

Given that the two systems are conditionally independent, one variable should not affect another on its trajectory, and so, those terms should not enter Jacobian entries. A consequence of this is that the $\eta\mu$ and $\mu\eta$ blocks partially comprising the upper and lower triangles of \eqref{jacobian} should vanish, i.e., each entry ought to be identically zero:
\[
\begin{bmatrix}
J_{\eta^1\eta^1} &\ldots &J_{\eta^1 b^1} & & & &\\
\vdots &\ddots &\vdots & & & 0 \\
J_{\eta^k\eta^1} &\ldots & J_{\eta^k b^1} &\ldots\\
J_{b^1 \eta^1} &\ldots & J_{b^1 b^1 } &\ldots &J_{b^1 \mu^1} &\ldots \\
\vdots &  &\vdots \\
 & & &\\
 & 0& & & & \ddots \\
 & & & & & & J_{\mu^m\mu^m}
\end{bmatrix}.
\]

In general, all the matrices indicated (excepting the Hessian) are matrix-valued forms or matrix fields, and hence we have denoted matrix multiplication by the more general wedge product, which reduces to a summed multiplication of smooth functions. Suppose instead the entries of $T$ are constant in states; in that case, we have $J = T H$, $H$ being the Hessian form. A zero entry in the Hessian form implies conditional independence of the statistics of two systems by arguments from information geometry, since those those two states carry no information about each other \cite{amari}, which is what we desire from the Markov blanket \cite{g-and-a}; more broadly, if any mixed partial derivative is zero, one (or both) of those states is a direction for which the system flows with at most a constant rate. In this way the state does not enter the system's dynamics usefully, and this recovers our original notion of the blanket between $\eta$ and $\mu$ \cite{stoch-chaos}. Therefore, we have 

\begin{conjecture}[sparse coupling]
Consider a class of systems where the flow in at least one direction is decoupled such that the system has a Markov blanket, i.e., where 
\[
J_{\eta^i\mu^j}J_{\mu^j\eta^i} = 0
\]
such that
\[
H_{\eta^i\mu^j} = 0
\]
for all $\eta\mu$ or $\mu\eta$ pairs, from which a Markov blanket follows by the argument above. All sufficiently high-dimensional, non-linear, coupled random dynamical systems that admit a stationary solution admit a stationary solution satisfying this property.
\end{conjecture}

In other words, this is a conjecture that conditional independence always follows from decoupling when in high-dimensions.

The inspiration for the sparse coupling conjecture is owed to the statistical physics of non-equilibrium systems. Physical systems with features of complexity are often high-dimensional and non-linear. As such, the sparse coupling conjecture is morally equivalent to stating that, in particular, physical systems which do not satisfy the indicated property comprise a set of measure zero with respect to those that do\textemdash and thus, that Markov blankets ought to be ubiquitous in nature under generic assumptions on what counts as a natural system.

We now refer to a critical identity derived by \cite{conor-comment} which is sufficient for the presence of a Markov blanket, by producing the relation sought in the sparse coupling conjecture. We will give a proof as a courtesy to the reader. A rephrasing of their result (for the original statement, see there) is as follows:

\begin{theorem}[Heins and Da Costa, 2022]\label{conor-thm}
Let $J = TH$ and let $T$ be linear in states. If the entry $J_{\eta^i\mu^j} = 0$, then the corresponding entry of $H_{\eta^i\mu^j}$ vanishes if and only if 
\begin{equation}\label{b-index-1}
\sum_{t\in\eta^i_C} Q_{\eta^i t} H_{t\mu^j} = 0.
\end{equation}
Alternatively, if $J_{\mu^j\eta^i} = 0$, then we have $H_{\eta^i\mu^j} = 0$ if and only if
\begin{equation}\label{b-index-2}
\sum_{t\in\mu^j_C} H_{\eta^i t} Q_{t \mu^j} = 0.
\end{equation}
\end{theorem}
\begin{proof}
Let $\eta^{i}_{C}$ be the complement of $\eta^i$, i.e., 
\[
\eta^i_C = \{\{\mu\}_u, \{b\}_v, \{\eta \setminus \eta^i\}_w\}
\]
with $u \in \{1, \ldots, m\}$ and so forth. Performing the matrix multiplication in \eqref{matrix-mult} we have 
\[
J_{\eta^i\mu^j} = T_{\eta^i\eta^i} H_{\eta^i\mu^j} + \sum_{t \in \eta^i_C} Q_{\eta^i t} H_{t\mu^j},
\]
for which some algebra yields
\[
H_{\eta^i\mu^j} = T_{\eta^i\eta^i}^{-1}\left(J_{\eta^i\mu^j} - \sum_{t \in \eta^i_C} Q_{\eta^i t} H_{t\mu^j}\right).
\]
Suppose $J_{\eta^i\mu^j} = 0$ and recall that $\Gamma$ is positive definite, so the diagonal entries $T_{\eta^i\eta^i}^{-1}$ are non-vanishing. Then $H_{\eta^i\mu^j}$ is identically zero if and only if
\[
\sum_{t \in \eta^i_C} Q_{\eta^i t} H_{t\mu^j} = 0.
\]
The case for $\mu^j\eta^i$ follows immediately.
\end{proof}

For clarity of computation we have denoted each multi-index and each sum over indices explicitly. We will refer to each of these quantities as a \emph{blanket index} as shorthand, with a formal definition to follow. This adapts the sparse coupling conjecture to a statement that for fairly generic random dynamical systems, and in particular, those relevant to statistical and biological physics, \eqref{b-index-1} or \eqref{b-index-2} (depending on the direction of flow) holds. Note that these identities concern a particular inner product of these matrix entries, the inner product of particular rows taken as vectors\textemdash taken as vectors which generate some flow in a particular direction of the reduced state space (i.e., modulo $\eta^i$), this has the geometric interpretation that sparsely coupled flows flow in orthogonal directions to influences from variables to which they are coupled, reflecting the decoupling in the Jacobian. In other words, controlled systems exist in lower-dimensional controlled submanifolds of their state space. 

It is obvious that the blanket index being generically zero in the space of systems implies the sparse coupling conjecture, since in that case, $J = 0$ implies $QH = 0$ for most systems; hence most systems with decoupled flows have Markov blankets. Nevertheless, it is clear from the nature of our blanket indices that sparse coupling is actually fairly stringent, in the sense that the flows in many different joint directions must be zero at once. This cannot be true in general\textemdash and we would not indeed want it to be. Most physical systems engage in coupling with other systems on some brief timescale and in some state variables; moreover, systems that do not are mathematically uninteresting.

Suppose instead that many, \emph{but not all}, of these complementary interactions are orthogonal. We could define this effective, but non-strict, sparse coupling as a \emph{weak Markov blanket}. 

\begin{defn}\label{b-index-def}
Suppose there are $m$-many $\mu$ components and $k$-many $\eta$ components. The blanket index of $J_{\eta^i\mu^j},$ denoted by $\Ind(J_{\eta^i\mu^j}),$ is \eqref{b-index-1} (likewise for $\Ind(J_{\mu^j\eta^i})$ and \eqref{b-index-2}). The blanket index of $J$ is 
\[
\sum_{i=1}^k \sum_{j=1}^m \left( \sum_{t\in\eta^i_C} Q_{\eta^i t} H_{t\mu^j} \right)
\]
if flows in the $\eta\mu$ direction are of interest, and likewise for the opposite case. This is the sum of every inner product we would need to consider, for whom a Markov blanket exists if and only if $\Ind(J)$ is identically zero and a weak Markov blanket exists only if $\Ind(J)$ is small. 
\end{defn}

The notion of a weak Markov blanket is thus a mere formalisation of the idea that systems that persist do so in the face of limited perturbations from the systems to which they are coupled, a particular sort of control. What qualifies this as a weak Markov blanket is that we do not demand that every degree of freedom on every timescale need be blanketed: we only need blankets on the ones we wish to measure. Critically, this is what it truly means for some statistic to be distinguishable. As such, we can have a system with key measurable properties that we desire to be statistically distinguishable on some time scale, which are blanketed, but in general it is not the case that the system will be perfectly sequestered from another system, neither physically nor statistically. Indeed, that case is equally vacuous: complex systems always have material turnover and exchange across a boundary. 

The physical \emph{desiderata} satisfied by stochastic systems with weak blankets motivates the following weakening of the sparse coupling conjecture:

\begin{conjecture}[non-strict sparse coupling]
Consider a class of systems where the flow in at least one direction is decoupled such that the system has a weak Markov blanket, i.e., where 
\[
J_{\eta^i\mu^j}J_{\mu^j\eta^i} = 0
\]
such that
\[
H_{\eta^i\mu^j} = 0
\]
for most $\eta\mu$ or $\mu\eta$ pairs, from which a weak Markov blanket follows by the argument above. All sufficiently high-dimensional, multiscale, coupled random dynamical systems that admit a stationary solution admit a stationary solution satisfying this property at some timescale. By consequence, physical systems which possess a steady state and do not satisfy this property anywhere at steady state comprise a set of measure zero with respect to those that do.
\end{conjecture}

The aim of this paper is now to prove the non-strict sparse coupling conjecture, and to see under what conditions we can also prove the original sparse coupling conjecture.

\section{The physics of weak Markov blankets in high-dimensional systems}\label{physics-sec}

We begin our results by establishing the validity of the weak Markov blanket approach. Here we show that the notions in the previous section make physical sense. We then establish that a consequence of the physical desire for weak blankets is an initially hypothesised sparsity in $Q$ which is useful for proving the sparse coupling conjecture. 

The following discussion makes reference to the adiabatic theorem. Broadly speaking, we have the following definition:

\begin{defn}
An adiabatic system is a system whose large-scale dynamics are invariant with respect to fast dynamical interactions.
\end{defn}

Consider a system where the interactions in $J$ come with a characteristic timescale. It is almost tautological that the measurable properties of a system remain stable and thus change on a much slower timescale than the fast fluctuations around it; this includes the material exchange of the system with an environment. Hence, adiabatically speaking, there exists a scale where a large number of zero $T$ terms exist due to this separation of timescales. The nature of this construction has already been discussed at length in \cite{afepfapp}. Instead of a formal discussion, we will sketch out why this argument makes sense and suggest that reference to the reader, the mathematical details of which (regarding separations of timescales especially) broadly follow our case.

Suppose there exist some essential set of states we wish to measure, on whose values the system's existence depends, such as order parameters in phased systems or control parameters in controlled systems. These have been referred to as existential variables in previous literature, and these are the variables we are interested in measuring, in the sense of the previous sections. Non-existential variables admit material exchange in the fashion of a weak Markov blanket; these are non-sparsely coupled states. However, being non-existential, this small degree of mixing does not appreciably dissipate one system into another. High-dimensional systems, having many more variables and many more time-scales, admit such weakenings without dissipating into an environment. This observation goes some way to formalising the above intuition: there are simply more places for a Markov blanket to be found; there is more room to not have strictly sparse coupling and still be effectively blanketed on some timescale for which those weakenings disappear adiabatically. This is why we observe coherent wholes in nature on average\textemdash because high-dimensional physical systems carve out submanifolds of their state spaces. 

On the other hand, it is also instructive where Markov blankets fail. Certain high-dimensional systems have strongly-mixing sub-components or sub-systems with long-range correlations that do exhibit extremely weak blankets, and which lead to a cohesive whole at a larger scale\textemdash a scale for whom \emph{those sub-systems} are the irrelevant small-scale objects. An example of such a construction is given in \cite{spin-glass}. Likewise, things situated in an environment that decay and mix into that environment take on the statistics of that environment, including its fast fluctuations, and are no longer identifiable as the stable phase they previously were in. Once again, this interplay is why there exist identifiable `things' in nature at every scale for which a particular notion of thing-ness is appropriate.

We conclude from this argument the following: in generic multiscale systems with many interacting degrees of freedom, the content of the adiabatic theorem is that there will be a timescale over which these fast fluctuations\textemdash or a spatial scale over which these short-range interactions\textemdash go away, leading to a scale over which most entries in $Q$ are naturally zero. By the adiabatic theorem this is the scale for which the system's dynamics remain invariant; simultaneously, by definition of a Markov blanket, this is the scale for which the system remains `system-like.' In probabilistic terms, this corresponds to a partial conditional independence, i.e., conditional independence in at least the subset of existential variables, or a non-essential degree of conditional dependence.\footnote{Recent literature in statistical physics \cite{knit} has provided evidence for the value of this approach by showing, numerically, that weakened blankets appear in non-equilibrium systems. \emph{Nota bene} the cited result is not an indictment of the concept of a Markov blanket. Rather, it begs exactly the question we have answered here.}

\section{The statistics of weak Markov blankets in high-dimensional systems}\label{quadratic-section}

In this section we justify our assumption of a quadratic $U$ in \eqref{sde}, based on an analysis of \eqref{mb-cond}. The main result here will support our contention that, if we are weakening the notion of a Markov blanket to essential conditional independence, then quadratic surprisal is sufficient to capture essential uninformativeness.

% Construct a manifold 
% \begin{defn}
% Let a probability density with a $\zeta$-dimensional vector of sufficient statistics, $\vartheta$, be parameterised by that vector. The moduli space of all such probability densities parameterised by all such values of $\vartheta$ is a $\zeta$-dimensional statistical manifold. 
% \end{defn}

Recall that a matrix of $\nu$-th derivatives of some function $f$ (here, denoted $\partial^\nu \!f$) is a $\nu$-tensor: if provided with $\nu$ vectors corresponding to $\nu$ directions of travel on some manifold, it returns a real number corresponding to the rate of change of $\nu$-dimensional volume forms on that manifold. For instance, `tiling' a manifold with two-forms (area elements) and finding the change in these tiles in some direction measures how they stretch or contract in that direction; indeed, the two-form corresponding to the matrix of second derivatives is known as the Hessian, and it does, in fact, measure curvature.

Here, the Hessian of surprisal measures how the surprisal of either state varies with the other state.\footnote{Throughout these results, we implicitly assume equality of mixed partial derivatives holds at least locally, placing minor restrictions on the regularity of our setting.} This is equivalent (in the very specific sense above) to asking how moving in the $(\eta^i, \mu^j)$ direction changes the surprisal.

\begin{lemma}\label{surprisal-lem}
If the Hessian of the surprisal in $(\eta^i, \mu^j)$ coordinates is zero, then there is no change in the surprisal when these two quantities are varied, or identically, when moving in the $\eta^i$ or $\mu^j$ direction: neither is informative to the other. Hence, the existence of a Markov blanket is implied. The converse also holds.
\end{lemma}
\begin{proof}
% Use the following facts: (i) $p(x \mid y \mid z) = p(x \mid y, z)$, and (ii) conditionalisation and marginalisation are well-defined operations; that is to say, they preserve identities. If $H_{\eta^i\mu^j} = \partial_{\eta^i}\partial_{\mu^j} \ln\{p(\eta^i, \mu^j)\} = 0$, then since $-\ln\{p(\eta^i \mid \mu^j) = -\ln\{p(\eta^i)\}$ (and vice-versa), for any variable $b$, we have $-\ln\{p(\eta^i \mid \mu^j, b) = -\ln\{p(\eta^i \mid b)\}$ (and vice-versa). Conversely, for any $b$ such that $-\ln\{p(\eta^i \mid \mu^j, b) = -\ln\{p(\eta^i \mid b)\}$, we have $-\ln\{p(\eta^i \mid \mu^j) = -\ln\{p(\eta^i)\}$. 
If 
\[
H_{\eta^i\mu^j} = -\partial_{\eta^i}\partial_{\mu^j} \ln\{p(\eta^i, \mu^j)\} = 0,
\]
then we have the following:
\[
-\partial_{\eta^i}\partial_{\mu^j}\big(\ln\{p(\eta^i \mid \mu^j)\} + \ln\{p(\mu^j)\} + \ln\{p(\beta)\}\big) = 0,
\]
where we introduce $\beta$ as an arbitrary state not equal to $\eta^i$ or $\mu^j$. Now conditionalise on $\beta$; then we have 
\begin{align*}
&-\partial_{\eta^i}\partial_{\mu^j}\big(\ln\{p(\eta^i \mid \mu^j \mid \beta)\} + \ln\{p(\mu^j \mid \beta)\} + \ln\{p(\beta \mid \beta)\}\big) \\ &\qquad = -\partial_{\eta^i}\partial_{\mu^j}\big(\ln\{p(\eta^i \mid \beta, \mu^j)\} + \ln\{p(\mu^j \mid \beta)\} + \ln\{p(\beta)\}\big) = 0.
\end{align*}
Assuming $\eta^i$ is not independent of $\mu^j$ on the nose, the above is zero if and only if 
\[
p(\eta^i \mid \beta, \mu^j) = p(\eta^i \mid \beta).
\]
Relabelling $\beta$ to $b$, we recover the Markov blanket we defined as the set of states for which $\mu^j$ is uninformative to $\eta^i$. The converse is more straightforward: if $b$ exists such that $p(\eta^i \mid b, \mu^j) = p(\eta^i \mid b)$, then the mixed partials indicated are zero. Both statements for the $\mu^j$ case follow identically.
\end{proof}

Now consider the case where higher-order surprisals are introduced, such that a zero in the two-tensor of surprisal does not imply a lack of dependence\textemdash for instance, a measurement of the change in volume forms 
\[
-\ln\{p(\eta^i, b, \mu^j, \mu^{\varsigma})\}.
\]
This is analogous to the (obviously trivial, but illustrative) one-dimensional case where $\partial_{\eta\eta\eta} \eta^3 = 6$, with $\partial_{\eta\eta}\eta^3 = 6\eta.$ When $\eta = 0$, the Hessian is zero, but the associated three-tensor is not. %We show this is vacuous. 
In this case we have the following:

\begin{prop}\label{nu-tensor}
Suppose the surprisal is a polynomial of degree $\nu = \mathrm{deg}(U)$ and both $m, k \geq \nu-1$. If pairwise conditional independence of $\eta^i$ and $\nu-1$ $\mu$'s implies $\eta^i$ is jointly conditionally independent of $\nu-1$ $\mu$'s, then there exists a strict Markov blanket for $\eta^i$ if and only if we have $\nu-1$ Hessian zeroes. The same holds for some $\mu^j$ and $\nu-1$ $\eta$'s.
\end{prop}
\begin{proof}
We will prove the claim by inducting on $\nu$. For an arbitrary direction $\mu^\varsigma$, let $\partial^3 U$ be $-\partial_{\eta^i\mu^j\mu^{\varsigma}} \ln\{p(\eta^i, b, \mu^j, \mu^{\varsigma})\},$ $\varsigma \neq j$. We have 
\[
-\partial_{\eta^i\mu^j\mu^{\varsigma}} \ln\{p(\eta^i, b, \mu^j, \mu^{\varsigma})\} = -\partial_{\eta^i\mu^j\mu^{\varsigma}} \ln\{p(\eta^i \mid b, \mu^j, \mu^{\varsigma}) p(\mu^j \mid b, \mu^\varsigma) p(\mu^\varsigma \mid b) p(b)\}.
\]
As such, a Markov blanket exists for $\eta^i$ when the surprisal has degree greater than two if and only if $\eta^i$ is jointly independent of $\mu^j$ and $\mu^\varsigma$ given $b$. We can use the probabilistic chain rule to show that the case for 
\[
-\partial_{\eta^i\mu^j\ldots\mu^{\varsigma}} \ln\{p(\eta^i, b, \mu^j, \ldots, \mu^{\varsigma})\}
\]
implies the case for 
\[
-\partial_{\eta^i\mu^j\ldots\mu^{\varsigma}\mu^\varrho} \ln\{p(\eta^i, b, \mu^j, \ldots, \mu^{\varsigma},\mu^\varrho)\},
\]
proving the statement by inducting on the extra coordinate. Now suppose we can conclude $p(\eta^i \mid b, \mu^j, \ldots, \mu^\varsigma) = p(\eta^i \mid b)$ if $p(\eta^i \mid b, \mu^j) = p(\eta^i \mid b)$ and $p(\eta^i \mid b, \mu^\varsigma) = p(\eta^i \mid b)$ and so forth for all other $\mu$ coordinates; in other words, that pairwise conditional independence is sufficient for joint conditional independence. Then we can split the condition on $\nu-1$ zeroes in $\eta^i$ entries of the $\nu$-tensor into $\nu-1$ zeroes in the $\eta^i$ entry of the Hessian. Then by Lemma \ref{surprisal-lem}, we have a (higher-order) Markov blanket between $\eta$ and those $\mu$'s. The direction of necessity is guaranteed since joint conditional independence implies pairwise conditional independence.
\end{proof}

\begin{remark}
Note that is is necessary but not sufficient that we simply have enough blanket states to reproduce pairwise independence\textemdash the key aspect of the above theorem is that pairwise independence, once achieved, implies joint independence, such that we may conclude a Markov blanket exists for higher-order tensors solely by evaluating the Hessian.
\end{remark}

%The logical inverse of this result is that if we cannot show pairwise conditional independence implies joint conditional independence, we must look at all orders of the $\nu$-tensor explicitly to verify there are no conditional dependencies.

We now have the following technical lemma and theorem:

\begin{notation*}
We denote $p(x \mid y, z) = p(x\mid y)$, i.e., that $x$ is conditionally independent of $z$ given $y$, by the shorthand $x \perp z \mid y$.
\end{notation*}

\begin{lemma}
If 
\begin{align*}
p(\eta \mid b, \mu^{\nu-1}) &= p(\eta \mid b) \\
p(\eta \mid b, \mu^{\nu-2}, \mu^{\nu-1}) &= p(\eta \mid b, \mu^{\nu-1}) \\
\ldots \\
p(\eta \mid b, \mu^1, \ldots, \mu^{\nu-1}) &= p(\eta \mid b, \mu^2, \ldots, \mu^{\nu-1})
\end{align*}
then 
\[
p(\eta \mid b, \mu^1, \ldots, \mu^{\nu-1}) = p(\eta \mid b).
\]
\end{lemma}
\begin{proof}
We will prove this inductively, beginning with the $\nu = 3$ case. Suppose $\eta \perp \mu^2 \mid b$ and $\eta \perp \mu^1 \mid b, \mu^2$. By the contraction principle, $\eta \perp \mu^1, \mu^2 \mid b$. Now pick an arbitrary $\nu$. Expanding the joint probability $p(\eta, b, \mu^1, \ldots, \mu^{\nu-1})$ into 
\[
p(\eta \mid b, \mu^1, \ldots, \mu^{\nu-1})p(\mu^1, \ldots, \mu^{\nu-1} \mid b)
\]
then if $\eta \perp \mu^1 \mid b, \mu^2, \ldots, \mu^{\nu-1}$ we have
\[
p(\eta \mid b, \mu^2, \ldots, \mu^{\nu-1})p(\mu^1, \ldots, \mu^{\nu-1} \mid b);
\]
if $\eta \perp \mu^2, \ldots, \mu^{\nu-1} \mid b$ also, then this equals
\[
p(\eta \mid b)p(\mu^1, \ldots, \mu^{\nu-1} \mid b),
\]
meaning $\eta \perp \mu^1, \ldots, \mu^{\nu-1} \mid b$. Now prove the same for $\eta \perp \mu^2, \ldots, \mu^{\nu-1} \mid b$: by the preceding argument, if $\eta \perp \mu^3, \ldots, \mu^{\nu-1} \mid b$ and $\eta \perp \mu^2 \mid b, \mu^3, \ldots, \mu^{\nu-1}$, we have $\eta \perp \mu^2, \ldots, \mu^{\nu-1} \mid b$. Shifting the first index we have the case for $\nu-2$ from the $\nu-1$ case. This closes the induction.
\end{proof}

We might call this the tower contraction lemma; it is a straightforward generalisation of the contraction principle.

\begin{theorem}
Let $s$ index $\{1, \ldots, \nu-1\}$. If all $\nu-1$ Hessian elements 
\[
-\partial_{\eta^i\mu^s} \ln\{p(\eta^i \mid b, \mu^1, \ldots, \mu^s)\}
\]
vanish, then there exists a Markov blanket for an arbitrary $\nu$-tensor applied to $\eta^i$ and $\nu-1$ $\mu$ directions.
\end{theorem}
\begin{proof}
To begin, note that 
\[
-\partial_{\eta^i\mu^s} \ln\{p(\eta^i \mid b, \mu^1, \ldots, \mu^s)\} = 0
\]
encodes the same blanket as
\[
-\partial_{\eta^i\mu^s} \ln\{p(\eta^i \mid b, \mu^s)\} = 0.
\]
That is, the extra conditions do not matter to the hypotheses of Proposition \ref{nu-tensor}. If these Hessian entries vanish, then by the tower contraction lemma, we have Proposition \ref{nu-tensor}. This proves the claim.
\end{proof}

Intuitively this is the statement that the curvature must be zero and the change in the curvature must be zero, but in the discrete limit, it is simply stating that the curvature in two non-identical directions are both zero. So, when we lose no information by passing to that limit, we can reduce one higher-order effect to multiple lower-order ones. Whilst this is not quite as trivial as it seems\textemdash as we descend the tower, there are more conditions for $\eta$ to be independent of\textemdash this means the quadratic surprisal is enough to capture the existence of a weak Markov blanket.

%edge cases where we have degeneracies? Can we have say 5th order surprusal but only eta1, eta2, mu1?

% ALSO this assumes we can convert p(eta | b, b', eta1, eta2) to p(eta | b, eta1) and p(eta | b', eta2) separately. is that possible?

\section{The presence of weak Markov blankets in high-dimensional systems}

We assume a state space globally isomorphic to $\R^n$ for ease, although the particularities of these proofs generalise to any sufficiently well-behaved metric space. Every such space is considered to be equipped with the Euclidean metric, i.e., that which is derived from the $\ell_2$ norm. Throughout we rely on the inductive statement that if a proposition holds for almost all objects (i.e., with high probability) at and above a sufficiently large $n$, and holds everywhere asymptotically, then it holds for almost all objects; clearly this is true, since we can prove infinitely more cases simply by taking $n\to\infty$. In some moral sense this is analogous to $\varepsilon$-regularisation, and this is precisely the motivation for our approach. 

Recall the statement of the sparse coupling conjecture. As intimated in the Introduction, there is a hint of ill-posedness to the conjecture: it does not ask to what extent a Markov blanket exists based on to what extent the system mixes with the environment. This makes it difficult to prove the conjecture formally, since we would need to explicitly construct a space of random dynamical systems and use some notion of the topology on that space to see how many such systems are perfectly blanketed.

However, in \cite{conor-comment}, we are provided with what is evidently an invariant for the classification of random dynamical systems: a quantity corresponding to any system's blanketed-ness, which here we assume corresponds to regions of the space of such systems which are more or less blanketed. Now we can simplify our approach to prove that this index is generically zero in the space of systems, and relate some measure on the region with blankets to some measure on that region without. 

We are now concerned with the relative blanketed-ness of a system, in terms of this index. Observe the following:

\begin{lemma}\label{ll-implies-NSCC}
Suppose all state variables are bounded above by some finite quantity $h$ and let $\Ind_{\emph{max}}(J)$ be the maximum possible blanket index of an assumed fully coupled system in dimension $n$,
\begin{align*}
\sum_{i=1}^k \sum_{j=1}^m h\big(1_{\eta^i \mu^1} 1_{\mu^1 \mu^j} &+ \ldots + 1_{\eta^i \mu^m}1_{\mu^m\mu^j} + 1_{\eta^i b^1} 1_{b^1 \mu^j} + \ldots + 1_{\eta^i b^l}1_{b^l\mu^j} \\ & \quad + 1_{\eta^i \eta^1}1_{\eta^1\mu^j} \ldots 1_{\eta^i\eta^{i-1}}1_{\eta^{i-1}\mu^j} + 1_{\eta^i\eta^{i+1}}1_{\eta^{i+1}\mu^j} + \ldots 1_{\eta^i\eta^{k}}1_{\eta^k\mu^j} \big) \\ {} \\ &\quad = \sum_{i=1}^k \sum_{j=1}^m h(k+l+m-1)_{ij} = kmh(k+l+m-1) = k m h (n-1).
\end{align*}
If the blanket index $\Ind(J)$ satisfies
\begin{equation}\label{norm-b-index-ll-1}
\abs{\frac{\Ind(J)}{\Ind_{\emph{max}}(J)}} \ll 1
\end{equation}
almost everywhere (i.e., for almost all $J$), then the non-strict sparse coupling conjecture is implied.
\end{lemma}
\begin{proof}
This follows from Theorem \ref{conor-thm}.
\end{proof}

We have previously argued that most entries in $Q$ will be zero for good physical reasons: in high-dimensional systems, not every internal degree of freedom will couple to every external degree of freedom on any appreciable timescale. The flow of some system is uniquely determined by some pair of $Q$ and $H$; as such, one can imagine drawing random samples from the space of $QH$'s, with each such sample being a different configuration of $Q$ with a different level of sparsity. A consequence of this physical argument is the \emph{ansatz} that the expected $Q$ will be zero. This allows us to use an argument from large deviations theory to treat realisations of possible $QH$'s as a random variable, and explicitly bound the probability of it being large (in particular, larger than the mean of zero) by an exponentially decreasing function. 

This leads us to an important set of hypotheses and then one of two main results:

\begin{remark}\label{hypotheses-rem}
We consider `a physical system,' to which the conjectures we are interested in apply, to be a multiscale system with the adiabatic characteristics investigated above. We further assume that $Q$ and $H$ are at most non-linearly correlated in most $t$, such that $\E[Q_{\eta^i t}H_{t\mu^j}] = \E[Q_{\eta^i t}]\E[H_{t\mu^j}]$ for most $t$, which is adiabatically zero. Finally, we assume that each $t$-th product, any possible $QH$ for different combinations of states, is independent when regarded as a random variable and is bounded above by some finite quantity $h$ (respectively, below, $-h$); the first assumption is justified by assuming in turn that there are no higher-order couplings between joint directions $(\eta^i t, t\mu^j)$ for different $t$, but is not necessary, whilst the second follows from finite-ness of the interaction energy of the system. It would be fair to say this is a more particular class of complex systems, rather than a class of generic physical systems; recent work \cite{karl-comment} has suggested that the way forward with the sparse coupling conjecture is precisely to qualify it from applying to `almost all physical systems' to `almost all sufficiently complex physical systems.' As we intimated in our inductive argument in the Introduction, there are many sorts of complex systems, of which there are many more than simple systems, recovering the spirit of the original statement of the conjecture.
\end{remark}

\begin{theorem}\label{main-result}
Suppose we have an $n$-dimensional adiabatic system linear in $T$ entries and arbitrarily distributed vectors $Q_{\eta^i}$ and $H_{\mu^j}$ satisfying the properties stipulated in Remark \ref{hypotheses-rem}. Let $\varepsilon > 0$ be some real number. When $n$ is large, we have a concentration of measure such that the probability of not possessing a Markov blanket in $\eta^i\mu^j$ coordinates falls off as $e^{-\frac{1}{2}(n-1)\varepsilon^2}$ for any lower bound on a partial blanket index, $\varepsilon$. The non-strict sparse coupling conjecture is implied and the sparse coupling conjecture holds asymptotically.
\end{theorem}
\begin{proof}
Take $\abs{\Ind_{\text{max}}(J_{\eta^i\mu^j})} = h(n-1)$ and let $X$ be the random variable 
\begin{equation}\label{def-of-X}
\frac{1}{h(n-1)}\sum_{t\in\eta^i_C} Q_{\eta^i t} H_{t \mu^j},
\end{equation}
our normalised blanket index from Proposition \ref{ll-implies-NSCC}. We established there that if this quantity is small the non-strict sparse coupling conjecture follows. Suppose $h$ bounds each term in $\Ind(J_{\eta^i\mu^j})$ such that, distributing the prefactor in \eqref{def-of-X} into the sum, we have
\[
\frac{Q_{\eta^it} H_{t\mu^j}}{h(n-1)} \in \tfrac{1}{n-1}[-1, 1].
\]
Suppose moreover that each $t$-th product of coefficients of $Q_{\eta^i}$ and $H_{\mu^j}$ is an independent random variable, with mean zero by hypothesis. We may apply the additive Chernoff bound to $X$, yielding 
\begin{equation}\label{chernoff-bound}
P\left(\abs{ X} > \varepsilon\right) < 2e^{-\frac{1}{2}(n-1)\varepsilon^2}
\end{equation}
for arbitrary $\varepsilon$. The union bound now yields
\[
P\left(\bigcup_{ij} \abs{X_{ij}} > \varepsilon\right) < \sum_i \sum_j 2e^{-\frac{1}{2}(n-1)\varepsilon^2} = 2kme^{-\frac{1}{2}(n-1)\varepsilon^2}.
\]
By Lemma \ref{ll-implies-NSCC}, we have the sparse coupling conjecture in the limit $n\to\infty$, and the non-strict sparse coupling conjecture elsewhere.
\end{proof}

For a review of tail estimates, see \cite{chernoff}. In some literature (including the cited paper) this particular form for the Chernoff bound is also called the Hoeffding inequality. A concise, if somewhat loose, summary of this result is as follows: the sparse coupling conjecture for certain $n$-dimensional systems holds with probability $1-2kme^{-\frac{1}{2}(n-1)\varepsilon^2}$. The probability of an $n$-dimensional $J$ being non-sparsely coupled in a way that does not satisfy either the weak or strong conjecture is small and vanishes as $n$ increases. As such, almost every system has a weak Markov blanket, and as dimension increases, Markov blankets strengthen with high probability. 

\begin{remark}
The asymptotic behaviour of \eqref{chernoff-bound} decreases so quickly in $n$ (relatively speaking) that we could potentially relax our assumption that $Q$ entries, and thus sums of products involving $Q$, are on average zero. In that case, large deviations from a small non-zero average would still be overwhelmingly unlikely, still rendering weak blankets common. We lose the consequential statement about the sparse coupling conjecture and are left only with non-strict sparse coupling.
\end{remark}

\begin{corollary}\label{part-dep-cor}
Theorem \ref{main-result} generalises to random $Q_{\eta^i t} H_{t \mu^j}$ partially dependent in $t$.
\end{corollary}
\begin{proof}
Following \cite[Theorem 2.1]{janson}, we relax the Chernoff bound to accommodate partial dependence. Let $\chi \geq 1$ be the smallest number for whom $\eta^i_C$ is covered by $\chi$ independent sets. In that case we have
\begin{equation}\label{part-dep-eq}
2\exp{-\frac{1}{2\chi}(n-1)\varepsilon^2}
\end{equation}
as the resulting Chernoff bound. If $\chi$ is small compared to $n$, i.e., if the variables are only partially dependent, then Theorem \ref{main-result} holds with bound \eqref{part-dep-eq}.
\end{proof}

This statement, an easy corollary of Theorem \ref{main-result}, covers cases where $Q_{\eta^i}$ and $H_{\mu^j}$ have a power series expansion in like variables\textemdash i.e., in coupling strength, as in \cite{aguilera2021}.\footnote{We thank Miguel Aguilera for suggesting the inclusion of a result in this direction.}

As a final note, the intuitive statement of Theorem \ref{main-result} is exactly the intuition that, in high-dimensional spaces, randomly chosen vectors are almost certain to be almost orthogonal; hence there is greater opportunity to have a weak Markov blanket. The large-deviations-type result establishes the commonality of weak Markov blankets in high-dimensions in probabilistic terms, complementing that geometric intuition. It is consistent with the well-known result that the normalised $\ell_2$ distance in high-dimensions is automatically sub-Gaussian with mean zero.

\section{Generalisation to the non-linear case}\label{non-lin-section}

The assumption that certain terms in the blanket index vanish only holds under certain linearity assumptions. Note the following:

\begin{prop}\label{b-index-nlin-prop}
The linear blanket index of Definition \ref{b-index-def} generalises to 
\[
\nInd(J_{ij}) = \sum_t Q_{it}(x) H_{tj} - \sum_t \partial_j T_{i t}(x) \partial_t U(x) + \sum_t \partial_{jt} T_{i t}(x)
\]
with $i \coloneqq \eta^i$ and $j \coloneqq \mu^j$.
\end{prop}
\begin{proof}
Let $x = (\{\eta\}, \{b\}, \{\mu\})$, indexed by $\alpha$. When the system is non-linear in $T$ entries, i.e., $T$ has a non-linear state-dependence, the Jacobian given in \eqref{jacobian} is
\[
J_{ij}(x) = \sum_t T_{it}(x) H_{tj} + \sum_t \partial_j T_{i t}(x) \partial_t U(x) - \sum_t \partial_{jt} T_{i t}(x).
\]
Rehearsing the argument from Theorem \ref{conor-thm}, we derive
\begin{align*}
\sum_t T_{it}(x) H_{tj} + \sum_t \partial_j &T_{i t}(x)\partial_t U(x) -\sum_t \partial_{jt} T_{i t}(x) = 0 \\
H_{ij} &= T_{ii}(x)^{-1}\left( \sum_t Q_{it}(x) H_{tj} - \sum_t \partial_j T_{i t}(x) \partial_t U(x) + \sum_t \partial_{jt} T_{i t}(x) \right),
\end{align*}
which is zero if and only if 
\begin{equation}\label{nlin-b-index}
\sum_t Q_{it}(x) H_{tj} = \sum_t \partial_j T_{i t}(x) \partial_t U(x) - \sum_t \partial_{jt} T_{i t}(x)
\end{equation}
for all $x^\alpha$.
\end{proof}

We caution the reader to note the greater stringency in the non-linear case; even though we have fixed $\eta^i$ and $\mu^j$, it is the case that \eqref{nlin-b-index} must be satisfied in all variables for $H_{ij}$ to be zero. From here onwards we will suppress the state-dependence of these matrices.

Denote by $\Phi$ the extra non-linear terms 
\[
\sum_t \partial_j T_{i t} \partial_t U - \sum_t \partial_{jt} T_{i t}.
\]
Na\"ively, as $n\to\infty$, we have $\tfrac{1}{n-1} \Phi \to 0$. This implies that, appropriately normalised, $QH = \Phi$ in the limit $n\to\infty$; thus, we have the sparse coupling conjecture by Proposition \ref{b-index-nlin-prop}. Let $h$ bound $Q_i H_j$ and $\Phi$ uniformly. To get a less trivial result we assume that $\Phi$ contracts only weakly, i.e., that $\Phi \propto n-1$ such that $\tfrac{1}{h(n-1)}\Phi \propto \frac{1}{h}$. The following theorem, a second and final key result, is possible even in this case:

\begin{theorem}\label{nlin-result}
Under the conditions assumed in Theorem \ref{main-result}, if 
\[
n-1+\varepsilon \leq \sum_t \partial_j T_{i t} \partial_t U - \sum_t \partial_{jt} T_{i t} \leq h,
\]
then Theorem \ref{main-result} generalises to the non-linear case.
\end{theorem}
\begin{proof}
Let $\sum_t \partial_j T_{i t} \partial_t U - \sum_t \partial_{jt} T_{i t}$ be denoted by $\Phi$, 
\[
\abs{\nInd_{\text{max}}(J_{ij})} = h(n-1)
\]
for some bound $h$ on every term in $(Q\wedge H)_{ij}$ and $\Phi$, and
\[
X = \frac{1}{h(n-1)}\sum_{t\in\eta^i_C} Q_{\eta^i t} H_{t \mu^j}.
\]
Now denote by $\Phi'$ our previous $\Phi$ normalised by $\nInd_{\text{max}}(J_{ij})$. Applying Chernoff's bound to $X$ relative to $\Phi'$, we have:
\begin{equation}\label{main-result-eq}
P\left(\Phi' + \varepsilon < \abs{X} < \Phi' - \varepsilon\right) < e^{-\frac{1}{2}(n-1)\frac{1}{(n-1)^2}\left(\Phi' + \varepsilon\right)^2} + e^{-\frac{1}{2}(n-1)\frac{1}{(n-1)^2}\left(\Phi' - \varepsilon\right)^2}.
\end{equation}
Since $(\Phi' \pm \varepsilon)^2 > 0$, the exponential decay of \eqref{main-result-eq} is guaranteed. The union bound applies in the obvious way. Take $\varepsilon$ to be a small, strictly positive real number. Now, for all $\Phi' \geq  n-1 + \varepsilon$, by Proposition \ref{b-index-nlin-prop}, the sparse coupling conjecture holds asymptotically; this generalises Theorem \ref{main-result}.
\end{proof}

\begin{corollary}
Theorem \ref{nlin-result} generalises to random $Q_{\eta^i t} H_{t \mu^j}$ partially dependent in $t$.
\end{corollary}
\begin{proof}
This follows the same argument as Corollary \ref{part-dep-cor}.
\end{proof}

We call attention to the particular fact that the extra terms\textemdash which one might expect to be difficult to control or to mix flows in different directions\textemdash actually allow for the same tail estimates, especially when they are large; when they are small, these estimates are better.

\begin{remark}\label{gaussian-remark}
Assuming non-quadratic surprisal, and thus non-constant $H$, one could shunt the state-dependence of $H$ into $Q$ and recover the case for constant $H$ by applying the theorem above. More generally, however\textemdash in the context of a Markov blanket, we can use Section \ref{quadratic-section} to claim we are only interested in conditional independence of pairs of variables; hence, we are only interested in zero Hessian entries, and do not care directly about the sorts of effects captured by higher-order derivatives of the surprisal. For this reason we can assume quadratic surprisal (and prove these results) without loss of generality. 
\end{remark} 

\section{Concluding remarks}

We have shown that, since the magnitudes of blanket indices scale inversely with dimension, a simple inductive argument holds that systems with more dimensions have smaller blanket indices; thus, that the cardinality of the set of systems with weak blankets can be taken to some large infinity. The set of low-dimensional systems for any appropriate finite cut-off cannot have larger measure than the set of systems for the set of high-dimensional systems on which the blanket index vanishes or becomes small. A restatement of the above result is that generic physical systems, especially those which are high-dimensional or multiscale, will possess weak Markov blankets. Hence Theorems \ref{main-result} and \ref{nlin-result} imply the non-strict sparse coupling conjecture. Since most every condensed-matter-type physical system is indeed multiscale, this is a result we can be certain of. At its physical core our results are nothing more than a formal statement that wholes in nature often consist of distinct parts and that systems which are broadly separated from their environments are separable in some broad sense.

We point out the following observation: our proof is only so elegant because we have ultimately proven relative blankted-ness, i.e., that any such system is comparatively weakly blanketed. This nonetheless implies non-mixture and a small blanket index given the `size' of the system; as such, this suffices to prove the sparse coupling conjecture.

The idea that strong Markov blankets are asymptotic constructions is consistent with the idea that Markov blankets are rare in simple systems \cite{aguilera2021} but ubiquitous in complex systems satisfying particular forms \cite{stoch-chaos, karl-comment}. It is obvious that simple objects have few states to couple sparsely, and thus have very sensitive Markov blankets: they either exist or they do not. The empirical observation that higher-dimensional and non-linear systems in physics are capable of engaging in more complex behaviours\textemdash and thus are better at cohesion and control\textemdash as well as semi-formal arguments about the ubiquity of sparsely coupled structures, all suggest that this can be anticipated in some generality. These statements lead to the perhaps non-obvious intuition that Markov blankets ought to be more commonplace in high-dimensional systems. The results here demonstrate this probabilistically, showing that samples of a large class of $Q$ will imply a Markov blanket with probability one asymptotically.

Complex systems are often high-dimensional and possess multiple scales; more generally, the intuitive statements of this result hinge on the physical fact that condensed matter and all systems at any macroscopic scale of observation necessarily consist of multiple scales. The fact that weak blankets are a general description of this aspect of complexity suggests their use is a promising approach to the study of complexity. Previous approaches built on the notion of a Markov blanket, like Bayesian mechanics and the free energy principle, are thus promising tools with which to study the mathematical physics of complex systems.

\bibliographystyle{alpha}
\bibliography{main}

\newcommand{\etalchar}[1]{$^{#1}$}
\begin{thebibliography}{DCFHP21}

\bibitem[Ama16]{amari}
Shun-ichi Amari.
\newblock {\em Information Geometry and its Applications}, volume 194 of {\em
  Applied Mathematical Sciences}.
\newblock Springer, 2016.

\bibitem[AMTB22]{aguilera2021}
Miguel Aguilera, Beren Millidge, Alexander Tschantz, and Christopher~L Buckley.
\newblock How particular is the physics of the free energy principle?
\newblock {\em Physics of Life Reviews}, 40:24--50, 2022.

\bibitem[APLHB22]{knit}
Miguel Aguilera, \'Angel Poc-L\'opez, Conor Heins, and Christopher~L Buckley.
\newblock Knitting a {M}arkov blanket is hard when you are out-of-equilibrium:
  two examples in canonical nonequilibrium models.
\newblock In {\em The Third International Workshop on Active Inference}, 2022.
\newblock Preprint arXiv:2207.12914. To appear.

\bibitem[Arn98]{arnoldRDS}
Ludwig Arnold.
\newblock {\em Random Dynamical Systems}.
\newblock Springer Monographs in Mathematics. Springer Berlin Heidelberg, 1998.

\bibitem[BTB{\etalchar{+}}21]{barp}
Alessandro Barp, So~Takao, Michael Betancourt, Alexis Arnaudon, and Mark
  Girolami.
\newblock A unifying and canonical description of measure-preserving
  diffusions.
\newblock 2021.
\newblock Preprint arXiv:2105.02845.

\bibitem[DCFHP21]{dacosta}
Lancelot Da~Costa, Karl~J Friston, Conor Heins, and Grigorios~A Pavliotis.
\newblock Bayesian mechanics for stationary processes.
\newblock {\em Proceedings of the Royal Society A}, 477(2256):20210518, 2021.

\bibitem[Doe20]{chernoff}
Benjamin Doerr.
\newblock Probabilistic tools for the analysis of randomized optimization
  heuristics.
\newblock In {\em Theory of Evolutionary Computation}, pages 1--87. Springer,
  2020.
\newblock See arXiv:1801.06733.

\bibitem[FHU{\etalchar{+}}21]{stoch-chaos}
Karl~J Friston, Conor Heins, Kai Ueltzh{\"o}ffer, Lancelot Da~Costa, and Thomas
  Parr.
\newblock Stochastic chaos and {M}arkov blankets.
\newblock {\em Entropy}, 23(9):1220, 2021.

\bibitem[Fri12]{friston12}
Karl~J Friston.
\newblock A free energy principle for biological systems.
\newblock {\em Entropy}, 14(11):2100--2121, 2012.

\bibitem[Fri19]{afepfapp}
Karl~J Friston.
\newblock A free energy principle for a particular physics.
\newblock 2019.
\newblock Preprint arXiv:1906.10184.

\bibitem[Fri22]{karl-comment}
Karl~J Friston.
\newblock Very particular: comment on {How Particular is the Physics of the
  Free Energy Principle?}
\newblock {\em Physics of Life Reviews}, 41:58--60, 2022.

\bibitem[HDC22]{conor-comment}
Conor Heins and Lancelot Da~Costa.
\newblock Sparse coupling and {M}arkov blankets: a comment on {``How Particular
  is the Physics of the Free Energy Principle?'' by Aguilera, Millidge,
  Tschantz and Buckley}.
\newblock {\em Physics of Life Reviews}, 42:33--39, 2022.

\bibitem[HKD{\etalchar{+}}22]{spin-glass}
Conor Heins, Brennan Klein, Daphne Demekas, Miguel Aguilera, and Christopher~L
  Buckley.
\newblock Spin glass systems as collective active inference.
\newblock In {\em The Third International Workshop on Active Inference}, 2022.
\newblock Preprint arXiv:2207.06970. To appear.

\bibitem[Jan04]{janson}
Svante Janson.
\newblock Large deviations for sums of partly dependent random variables.
\newblock {\em Random Structures \& Algorithms}, 24(3):234--248, 2004.

\bibitem[Pea88]{pearl}
Judea Pearl.
\newblock {\em Probabilistic Reasoning in Intelligent Systems: Networks of
  Plausible Inference}.
\newblock Morgan Kaufmann, 1988.

\bibitem[Sak22a]{g-and-a}
Dalton A~R Sakthivadivel.
\newblock Towards a geometry and analysis for {B}ayesian mechanics.
\newblock 2022.
\newblock Preprint arXiv:2204.11900.

\bibitem[Sak22b]{classical-physics}
Dalton A~R Sakthivadivel.
\newblock A worked example of the {B}ayesian mechanics of classical objects.
\newblock In {\em The Third International Workshop on Active Inference}, 2022.
\newblock Preprint arXiv:2206.12996. To appear.

\end{thebibliography}

\end{document}